# Influence of applied stress on energy dissipation and crack growth in articular cartilage

Dipul Chawla[1], Eric Kazyak[1], Melih Eriten[1], and Corinne R. Henak[1,2,3]

[1]Department of Mechanical Engineering, University of Wisconsin-Madison, Madison, WI
[2]Department of Biomedical Engineering, University of Wisconsin-Madison, Madison, WI
[3]Department of Orthopedics and Rehabilitation, University of Wisconsin-Madison, Madison, WI

*Address Correspondence to:
Corinne R. Henak, PhD
Assistant Professor
3031 Mechanical Engineering Building
1513 University Ave
Madison, WI 53706
608-263-1619
chenak@wisc.edu

**Graphical Abstract**

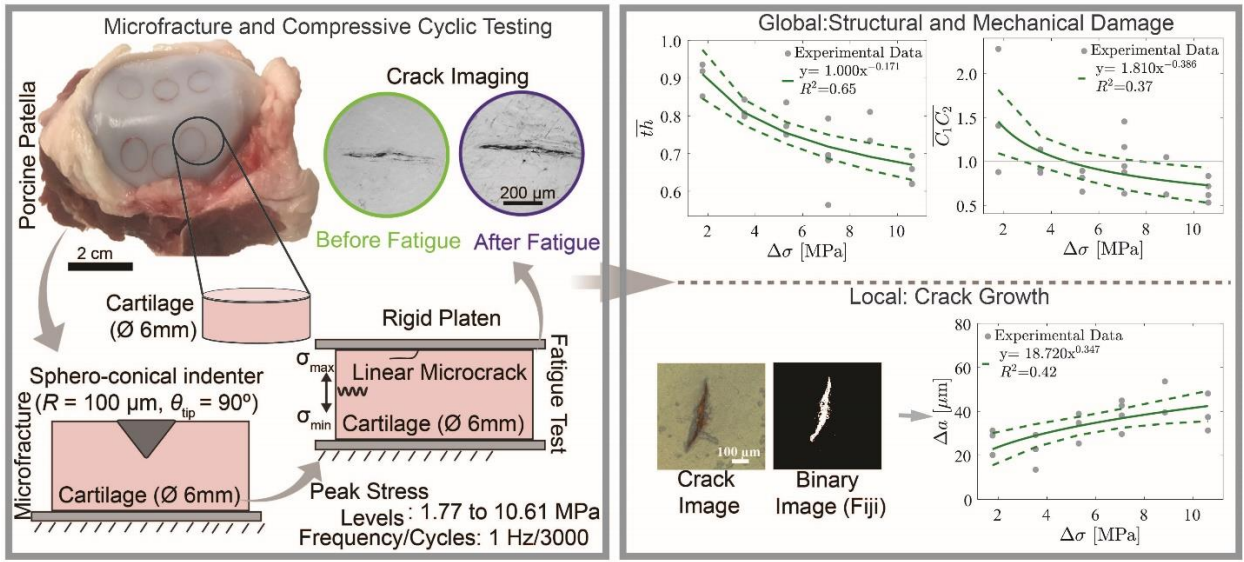

**Abstract**

Mechanical stress-induced damage to the articular cartilage can result in fracture imitation or significant tissue degeneration leading to osteoarthritis (OA) compromising joint mobility. Despite the clinical significance, a comprehensive understanding of the crack progression in cartilage damage remains elusive due to complex mechanical responses and variable initial crack geometry. This study investigated the impact of physiologically relevant stress levels, ranging from light-weight to high stress activities, on crack propagation in cartilage under compressive cyclic loading. Through empirical analysis, interconnections between localized damage (crack growth) and global damage (changes in material properties) across various levels of applied stress were established. Microscale cracks were nucleated in cylindrical cartilage plugs extracted from porcine patella specimens *via* microindentation and then subjected to controlled cyclic loading to examine changes in structural integrity, mechanical properties, and crack extension. Results demonstrated a decrease in cartilage thickness and apparent stiffness with increasing stress levels, indicative of bulk material damage. Dynamic mechanical properties such as energy dissipation and phase angle showed an overall reduction after fatigue loading. Crack growth followed an empirical law with increasing applied stress, reflective of greater mechanical damage at higher stress values. The findings of the study showed the interplay between global and local damage in cartilage under cyclic loading, suggesting insights into the fundamental changes occurring within the tissue. The outcomes of this investigation contributed to detailed understanding of the cartilage fatigue behavior through empirical laws and could potentially serve to prevent or delay tissue degeneration in OA. The findings could additionally be used for developing bio-inspired engineering materials for various applications.

# 1. Introduction

Cartilage damage includes local damage in the form of fissures and cracks, and global damage in the form of softening or decreased energy dissipative capacity, which can progress into osteoarthritis (OA) (Meachim, 1972). Because cartilage has limited healing capacity (Buckwalter, 1987), understanding the progression of damage is central to understanding the OA disease process. Cartilage is poroviscoelastic (PVE) (Chiravarambath et al., 2009; Edelsten et al., 2010; Lakes, 2009; Lawless et al., 2017; Mow et al., 1992), therefore, its mechanical and failure responses are expected to depend on both loading magnitude and rate. Damage may result in visible cracks, or diffuse damage to the extracellular matrix components, type II collagen and proteoglycans (PGs) with negatively charged glycosaminoglycan (GAG) side chains.

Prior evaluation of damage progression under cyclic loading in cartilage indicates that rate, stress amplitude, and number of cycles influence damage progression, yet there are no empirical laws for describing that damage progression. Previous research has demonstrated increased crack length with increased frequency (ranging from normal to rapid heal strike) (Chawla et al., 2024; Sadeghi et al., 2018a, 2015) and number of cycles (Sadeghi et al., 2018b). Global changes in the mechanical response of cartilage following dynamic loading include reduced dynamic modulus and phase angle (Park et al., 2004a), with damage resulting from collagen fibril disruption and PG aggregation under different levels of stresses mimicking light-weight (Morrison, 1970a, 1970b), high impact (Brown and Shaw, 1983; Smidt, 1973), and strenuous activities (Clements et al., 2001; Hodge et al., 1986; Matthews et al., 1977). Reduced cartilage thickness following fatigue loading has also been observed (McCormack and Mansour, 1998; Weightman, 1976). However, these studies have not reported relationships that can be used to understand crack extension or structural damage as a function of stress level under compressive loading.

In contrast, crack propagation mechanisms have been explored both mechanistically and empirically for engineered materials, typically relating an incremental increase in crack length ($da$) to stress level or stress intensity factor. Frost *et al.* (Frost and Dugdale, 1958) postulated that the rate of crack growth correlated with applied stress (obtained by fitting experimental data) and the initial crack length. Liu *et al.* considered an ideal elastic-plastic stress-strain profile and included total energy absorbed up to failure (Liu, 1961). McEvily and Illg proposed crack extension per cycle number to vary as function of $\sigma_o$, where $\sigma_o$ is the product of elastic-stress concentration factor ($K_N$) and net stress ($\sigma_{net}$) at the crack tip of radius $\rho$ (McEvily Jr and Illg, 1958). Alternatively, Paris and Erdogan postulated that the rate of crack growth correlated with the stress intensity factor via a power-law relationship (Paris and Erdogan, 1963). These relationships effectively describe crack extension in engineering materials with standardized crack-propagation configurations. Given the assumptions built into their derivations, though, these laws have limitations even for engineering materials. Notably, Paris and Erdogan concluded that "laws which correlate a wide range of test data from many specimens are perhaps the 'correct' laws" (Paris and Erdogan, 1963), implying the need to collect and fit experimental data in order to generate laws that hold for damage progression for complex materials such as articular cartilage.

Therefore, the main objective of this study is to empirically establish relationships between local damage (crack growth) and applied stress, as well as global damage (changes in material properties) and applied stress during fatigue loading. We aim to confirm greater crack extension with greater applied stress magnitude ($\Delta\sigma$) in cartilage under compressive cyclic loading. We hypothesize greater decreases in thickness (*th*), energy dissipation (*ED*), apparent stiffness (Veronda-Westmann coefficient, $C_1C_2$), and phase angle ($\phi$) with increase in stress levels.

## 2. Methods

*2.1 Specimen preparation*

Eighteen full thickness, 6-mm diameter cylindrical plugs were obtained from porcine patellae (5-6 months old, 8 animals, 1-5 samples per animal, sex unknown and assumed random) using biopsy punches and surgical blades. Joints were harvested from a local abattoir and were stored at -20º C wrapped in gauze soaked with Dulbecco's phosphate-buffered saline (DPBS) until dissected. Freeze-thaw cycles were kept to the minimum possible ($n=3$) to avoid significant changes in sample properties (Qu et al., 2014). Subchondral bone was removed using a razor blade and custom-made 3D printed fixture (Chawla et al., 2021). Before mechanical testing, sample thickness was measured at 3 locations with digital calipers (accuracy: ±0.02 mm, General Ultratech, General Tools & Instruments LLC., Secaucus, NJ) and the average thickness (*th*) was calculated. The bottom surface of the sample was securely attached to the fixture using cyanoacrylate (Scotch 3M super glue, 3M, Minnesota). All experimental testing, crack nucleation (*section 2.2*) and cyclic loading (*section 2.3*), was performed on a tabletop test machine (TA ElectroForce 3230-AT Series III, TA Instruments, New Castle, DE). Samples were kept in DPBS throughout testing to maintain hydration.

*2.2 Nucleation of microcracks*

Microindentation tests were performed to nucleate surface cracks as previously described (Chawla et al., 2022, 2021). Briefly, displacement-controlled tests were conducted using a diamond coated sphero-conical indenter with a tip radius, $R = 100$ µm and tip angle, $\theta_{tip} = 90º$ (Anton Paar, Austria) (Figure 1A). Samples were indented at 5 mm·s$^{-1}$ to 250 µm while the load was measured using a load cell (22 N capacity with 0.02 N resolution). Displacement and force data were acquired at

1700 Hz. The displacement rate was selected to obtain consistent linear cracks with comparable lengths (*section 2.4*).

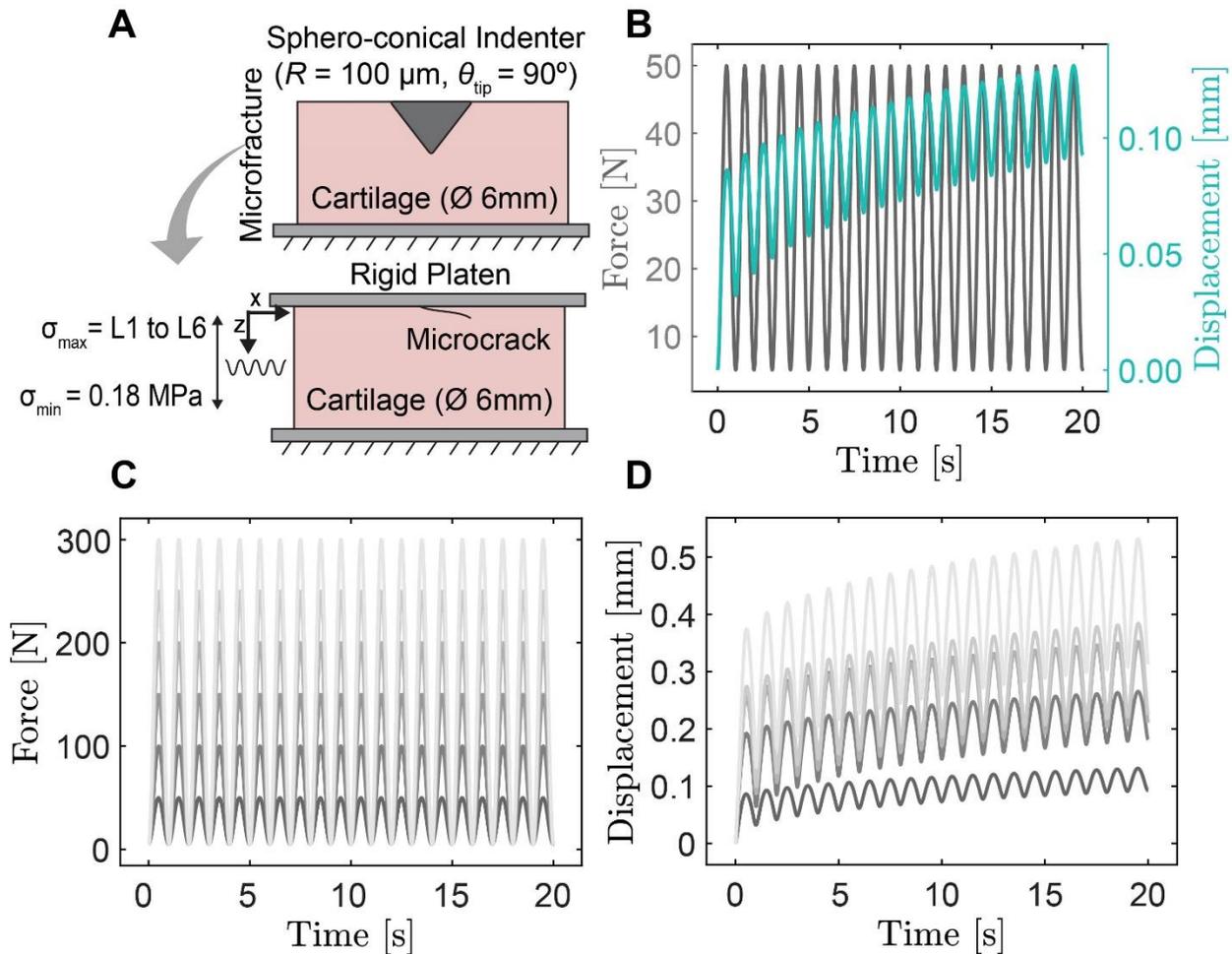

**Figure 1.** Schematic of the experimental setup and representative loading waveforms. (A) Consistent linear cracks were nucleated using microfracture tests with a sphero-conical tip and propagated using force-controlled cyclic loading. (B) Representative load (left axis) and displacement (right axis) *vs* time for pre-and post-diagnostics load-controlled testing. There was a 7-minute recovery period (between pre-diagnostics and fatigue loading) and 15-minute recovery period (between fatigue loading and start of post-diagnostics). (C) Representative load *vs* time data (first 20-cycles) from different levels of load-controlled fatigue tests (maximum load ranging from 50 N to 300 N). (D) Representative displacement *vs* time response (first 20-cycles) based on levels of maximum load or stress applied in fatigue tests.

*2.3 Load-controlled cyclic tests*

Microfractured cartilage samples were subjected to load-controlled cyclic sinusoidal waveform to cause changes in the mechanical properties and crack morphology (Figure 1A). Samples were loaded to six levels of maximum stresses matching physiological stress levels, ranging from lowest level (light-weight activities (Morrison, 1970b; Park et al., 2004b)) at L1=50 N and first Piola-Kirchhoff stress $\sigma_{max} = 1.77$ MPa to the highest stress level range (strenuous activities (Hodge et al., 1986; Matthews et al., 1977)) at L6=300 N and first Piola-Kirchhoff stress $\sigma_{max} = 10.61$ MPa. The intermediate stresses had an increment of 50 N, $\sigma_{max} = 1.77$ MPa per level. The minimum value of load and first Piola-Kirchhoff stress to keep the platen in contact was 5 N corresponding to $\sigma_{min} = 0.18$ MPa. All samples were loaded at 1 Hz (reflecting walking frequency (Simon et al., 1981)) for 3000 cycles. The cycle number was chosen based on preliminary studies that demonstrated measurable crack extension after 3000 cycles.

Before and after cyclic loading, samples underwent diagnostic testing to evaluate changes in material properties. Samples were loaded to 50 N at 1 Hz for 20 cycles (defined as pre/post diagnostics in this study) using a flat platen (Figure 1B). Samples were unloaded and left in DPBS for 7 minutes (between pre-diagnostics and cyclic loading) and for 15 minutes (between the end of cyclic loading and post-diagnostics) to reach equilibrium. Our prior publication demonstrated that that sample thickness did not change after the selected recovery times (Chawla et al., 2024). Displacement, time, and force response were recorded throughout testing using a 450 N load sensor (0.05 N resolution). All time series data per cycle were obtained by splitting the full data in time-intervals of 1 seconds (frequency = 1 Hz) with representative data for 20 cycles shown in Figure 1C-D. Tissue thickness prior to post-diagnostics ($th_{post}$) was determined by reading the actuator position at the point where the load cell detected minimum contact force (5 N) at the interface between platen and cartilage. The total irreversible compaction of the cartilage caused by fatigue loading is reported as normalized thickness, $\overline{th} = th_{post}/th$ for the remainder of the manuscript.

Changes in the mechanical behavior under different stress levels of cyclic loading were assessed from the variation between pre-and post-diagnostics (expressed as normalized parameters, i.e. $^{post}/_{pre}$). Engineering strain, stretch ratio ($\lambda$), 1st Piola-Kirchoff stress, and Cauchy stress (assuming isotropic behavior, incompressibility, and approximating the stress to uniaxial in the depth direction) were calculated from displacement, force, original sample area, and the reference thickness before the start of the test. These data were obtained from the loading part of the 1st cycle (pre-and post-diagnostics). Data were fit to an incompressible Veronda-Westmann strain-energy density function (Maas et al., 2012; Veronda and Westmann, 1970) (Eq. (1)).

$$W = C_1\left(e^{C_2(\tilde{I}_1-3)} - 1\right) - \frac{C_1 C_2 (\tilde{I}_2-3)}{2} \qquad \text{Eq. (1)}$$

$$T_{33} = 2C_1 C_2 e^{C_2(\lambda^2+\frac{2}{\lambda}-3)}[\lambda^2 - \frac{1}{\lambda}] + C_1 C_2 \left(\frac{1}{\lambda^2} - \lambda\right) \qquad \text{Eq. (2)}$$

Where $\tilde{I}_1$ and $\tilde{I}_2$ are the deviatoric components of the first and second invariants of the left stretch tensor, $B$, $T_{33}$ is the Cauchy stress along the loading axis (Eq. (2)), and $C_1$ and $C_2$ are the material parameters of the Veronda-Westmann constitutive model. The product of the two materials constants, $C_1 C_2$, is referred to as *apparent stiffness* in this study. To confirm that our observed trends were not a result of the assumptions made (incompressibility, uniaxial stress), we also fit the 1st Piola-Kirchoff stress versus engineering strain of the loading region of the first cycle using a mathematical (phenomenological) equation (Eq. (3)).

$$\sigma = A\left(e^{B*\varepsilon} - 1\right) \qquad \text{Eq. (3)}$$

Where $\sigma$ is the 1st Piola-Kirchoff stress, $\varepsilon$ is the engineering strain and $A$ and $B$ are the material parameters of the mathematical model. Both models fit the data well (Figure 2) and resulted in the same trends with loading levels (Figure S2). For conciseness, only the Veronda-Westmann material constants will be used for the remainder of the manuscript.

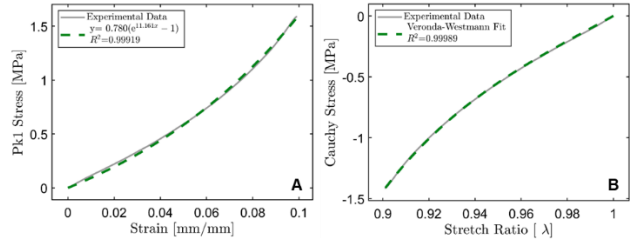

**Figure 2.** Comparison between mathematical model and constitutive model. (A) mathematical model fit (Eq. 3) to 1$^{st}$ Piola-Kirchoff (PK1) stress magnitude and engineering strain magnitude of the loading region of the first cycle (pre-diagnostics) of a representative sample. (B) Veronda-Westmann constitutive model fit (Eq. 2) to Cauchy stress and the stretch ratio of the loading region of the first cycle (pre-diagnostics) of a representative sample.

Energy dissipation was determined by calculating the area between the loading and unloading curves using trapezoidal integration. Cycles 17 through 20 were averaged for energy dissipation and phase shift calculation. Phase angle ($\phi$), was evaluated by comparing the phase differences (Eq. (4)) after fitting strain and stress data to sinusoidal functions of stress (Eq. (5)) and strain (Eq. (6)) where $\sigma_o$ and $\varepsilon_o$ represents maximum amplitudes, $\omega$ indicates angular frequency, $\phi_f$ and $\phi_d$ denotes phase differences in sinusoidal functions with experimental data, and *t* is the time. All the calculations were made using custom MATLAB scripts (version 2021a, Mathworks, Inc., Natick, MA).

$$\phi = \phi_f - \phi_d \qquad \text{Eq. (4)}$$

$$\sigma(t) = \sigma_o \sin(\omega t + \phi_f) \qquad \text{Eq. (5)}$$

$$\varepsilon(t) = \varepsilon_o \sin(\omega t + \phi_d) \qquad \text{Eq. (6)}$$

*2.4 Measurement of crack length*

Microfractures were assessed through imaging. India ink was applied to the articular surface prior to crack nucleation (before pre-diagnostics) and after cyclic loading tests (after post-diagnostics). Excess ink was gently wiped off with a DPBS dampened wipe (Kimtech, Orange, TX). Sample surfaces were imaged using a LEXT OLS5100 3D Measuring Laser Microscope (Evident Corporation, Tokyo, Japan) with a 10× objective. The images were processed with "manual-tilt correction" using post-processing software (LEXT analysis application). An overlay of the crack morphology (Figure 3B) was obtained by averaging three images: white light, intensity map from 405 nm blue laser, and height (*z*) profile. Semi-automated crack length measurements were made using a custom macro in Fiji (ImageJ version 1.54g). All crack images corresponding to respective stress levels are shown in the supplemental information (Figures S5-S10).

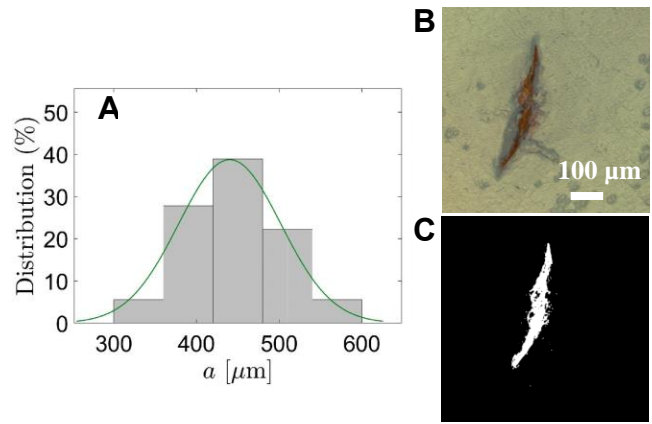

**Figure 3.** Initial crack length distribution and representative measurement methodology. (A) histogram of the initial crack length (*a*). (B) Representative crack morphology by averaging three images: white light, height profile, and intensity map from 405 nm blue light. (C) Representative binary image processed in Fiji used for crack length measurement.

*2.5 Statistical analysis*

To investigate the possible structure-mechanical function relationships, power-law based regression analyses were performed on the normalized variables ($\overline{th}$, $\overline{C_2}$, $\overline{C_1 C_2}$, $\overline{\phi}$, and $\overline{ED}$) as function of applied stress amplitude ($\Delta\sigma = (\sigma_{max} - \sigma_{min})$). To analyze crack growth behavior with different physiological stress levels, power-law regression analysis was performed on the change in the crack length ($\Delta a$) as a function of stress level ($\Delta\sigma$). All regression analyses and curve fitting were conducted in MATLAB (The MathWorks, Inc., Natick, MA).

## 3. Results

*3.1 Material stiffness and compaction after cyclic loading*

Cartilage compaction under cyclic (fatigue) loading induced structural damage with normalized thickness, $\overline{th}$ decreasing with increase in stress level (Figure 4). $\overline{th}$ decreased by 27.1% following a power-law relationship, $y = 1.004x^{-0.171}$ with $R^2$=0.65, from 0.91 ± 0.04 at the lowest stress level (L1) to 0.66 ± 0.03 at the highest stress level (L6).

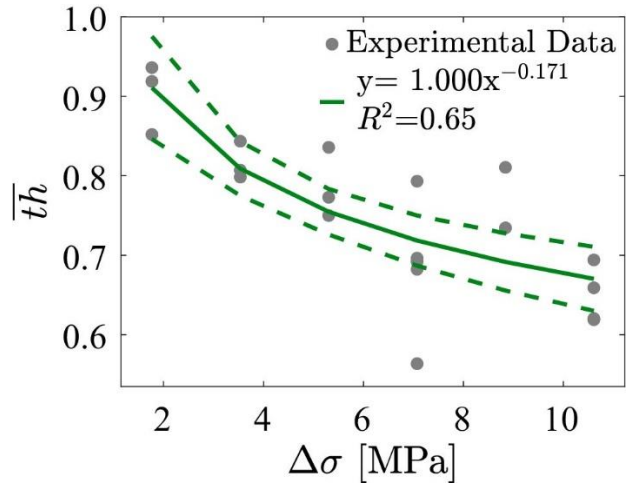

**Figure 4.** Effect of stress levels on structural integrity. Normalized thickness $\overline{th}$ decreased after cyclic loading following a power law with increase in physiological stress level ($\Delta\sigma$). Dashed lines show the 95% confidence intervals.

Changes in the material constants ($C_1C_2$ and $C_2$) indicated structural damage after fatigue loading (shown as expressed as normalized variables in Figure 5). $\overline{C_1C_2}$ decreased by 55.2% with a power-law trend, $y = 1.810x^{-0.386}$ and $R^2$=0.37, from 1.52 ± 0.71 at L1 stress level (1.77 MPa) to 0.68 ± 0.13 at L6 stress level (10.61 MPa) (Figure 5B). Similarly, $\overline{C_2}$ decreased with a power-law trend, $y = 8.840x^{-0.918}$ and $R^2$=0.37, varying by 74.6% from 5.48 ± 4.83 at L1 stress level (1.77 MPa) to 1.39 ± 0.71 at L6 stress level (10.61 MPa) (Figure 5A).

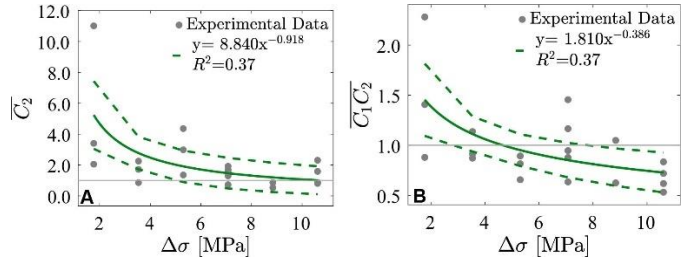

**Figure 5.** Effect of stress levels on mechanical behavior. (A) Normalized nonlinear second Veronda-Westmann material constant $\overline{C_2}$ and (B) Normalized apparent stiffness $\overline{C_1 C_2}$ decreased after loading following a power law with an increase in stress level (Δσ) indicating cartilage softening at higher stress levels (L4=7.07 MPa to L6=10.61 MPa). Dashed lines show the 95% confidence intervals.

*3.2 Phase angle and energy dissipation before-and after fatigue loading*

Both phase angle, $\phi$ and energy dissipation, *ED* decreased after fatigue (expressed as normalized variables in Figure 6). However, there was no correlation found between $\bar{\phi}$ or $\overline{ED}$, and stress level.

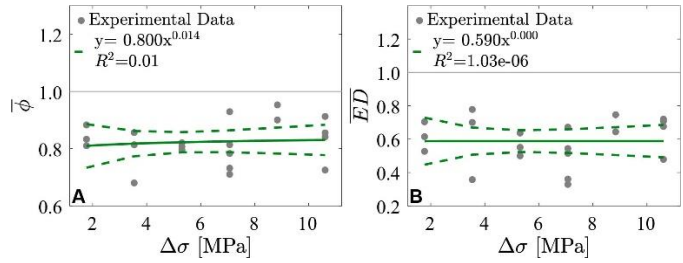

**Figure 6.** Effect of stress levels on dynamic mechanical behavior under compressive cyclic loading. (A) Normalized phase angle $\bar{\phi}$ and (B) Normalized energy dissipation ratio $\overline{ED}$ were not correlated with stress levels (Δσ). Dashed lines show the 95% confidence intervals.

*3.3 Correlation between change in crack length vs stress level*

Microfractured cartilage samples had consistent linear cracks (Figure 3, Figures S5-S10) that extended under fatigue loading (Figure 7). Crack growth increased with increased stress levels following a trend $y = 18.720x^{0.347}$ and $R^2$=0.42 ranging from 26.82 ± 4.79 µm at the lowest stress level to 38.96 ± 6.95 µm at the highest stress level (Figure 7). Variation in the normalized change in crack length; Δ*a/a* also followed a power law (Figure S4).

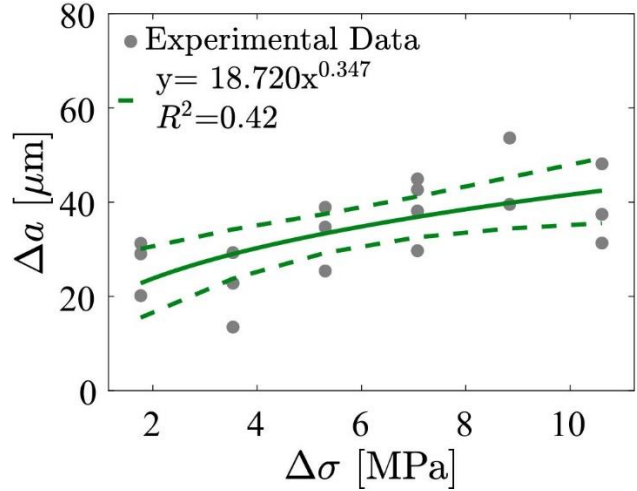

**Figure 7.** Crack growth after compressive cyclic (fatigue) loading. Change in crack length, $\Delta a$, followed a power with increase in stress level ($\Delta \sigma$). Dashed lines show the 95% confidence intervals.

## 4. Discussion

*4.1 Establishment of empirical laws to describe cartilage fatigue damage*

This work postulates power-law based empirical relationships between structural damage (representing bulk material damage, detailed in *section 4.2*) and crack growth (localized damage, detailed in *section 4.3*) as a function of physiologically-relevant applied stress levels. In the realm of bulk material response, $\overline{th}$, decreased following $y = 1.004x^{-0.171}$, $R^2$=0.65 with increase in stress level indicating overall irreversible compaction. The experimental data showed decreased normalized Veronda-Westman material constants ($\overline{C_2}$ and $\overline{C_1 C_2}$) with increased stress levels following $y = 8.840x^{-0.918}$, $R^2$=0.37 and $y = 1.810x^{-0.386}$, $R^2$=0.37, respectively, suggesting bulk softening of cartilage after post-diagnostics (at higher levels from L4=7.07 MPa to L6=10.61 MPa) under compressive cyclic loading. Crack length, $\Delta a$ increased following $y = 18.720x^{0.347}$, $R^2$=0.42 suggesting more pronounced localized damaged at elevated stress levels. Although the correlation is weak, the observed change in crack length, ($\Delta a$) are in line with prior studies that demonstrated greater crack growth with increase in strain-energy release rate ($G$) (Weizheng Li et al., 2022; Ni et al., 2020; Scetta et al., 2021; Xu et al., 2023). The empirical laws fitted to the experimental data in this study showed high resistance to increased crack length, ($\Delta a$) even at elevated levels of stress levels with a power law constant of 0.347. In this study, we focused on investigating the correlation between the change in crack length ($\Delta a$) with stress levels to understand localized damage as compared use of strain-energy release rate ($G$) or stress-intensity factor ($k$) in prior studies on standard engineering materials (e.g., (Frost and Dugdale, 1958; Head, 1953; Liu, 1961; McEvily Jr and Illg, 1958; Paris and Erdogan, 1963)), so the constants are not directly comparable.

*4.2 Global damage after cyclic loading: Structural and mechanical changes*

Structural damage and alterations in material properties were load-level dependent and are postulated to reflect changes in microstructure (Figure 8). Cartilage thickness (*th*) decreased from before-to-after testing across conditions and normalized cartilage thickness ($\overline{th}$) decreased with increasing load levels (Figure 5). This is consistent with previous studies (Clements et al., 2001; Hosseini et al., 2014; Kaplan et al., 2017; Lyyra et al., 1995; Neu et al., 2005) that have shown compromised extracellular matrix, physical damage to microstructure including PGs, GAGs, and collagen damage. Prior research has demonstrated differences in the recovery strain in intact *vs* surface removed cartilage by applying static and cyclic creep loading (Torzilli and Allen, 2022). However, we uncoupled the recoverable strain from changes in thickness due to damage by allowing cartilage to recover prior to taking post-fatigue measurements and therefore consider the changes in thickness as irrecoverable damage. While this study did not assess microstructural composition, the reduction in normalized thickness (post/pre-diagnostics) implies irreversible structural damage to the *ex-vivo* tested cartilage.

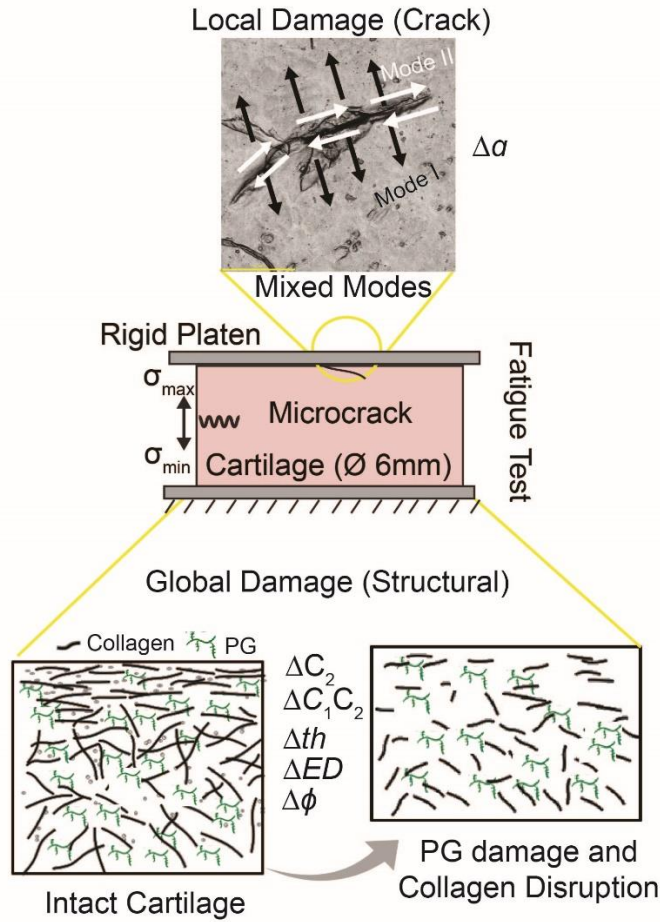

**Figure 8.** Local damage (crack growth) and global damage (changes in structural-mechanical properties) follows compressive cyclic (fatigue) loading.

Cartilage material properties ($\overline{C_2}$ and $\overline{C_1 C_2}$) increased under low load levels but decreased under high load levels (Figure 5), indicating interplay between compaction, apparent stiffness ($C_1 C_2$) and nonlinearity ($C_2$). These trends are also consistent with prior work indicating that low load magnitude (20 N) resulted in negligible collagen network damage compared to severe collagen damage at the highest loading magnitude (200 N) (Kaplan et al., 2017). Thus, the decreased thickness but increased stiffness measured at lower stresses (L1=1.77 MPa to L3=5.31 MPa) in this study may indicate alterations in the PG matrix but not the collagen fibril matrix. The increase in apparent stiffness at low stress magnitudes could be attributed to altered electrostatic repulsion (Guterl et al., 2010; Jin and Grodzinsky, 2001; Lu et al., 2023) between PGs (negatively charged GAGs) with negligible collagen damage.

The results shown here for cartilage can be compared to those from synthetic soft materials. Prior studies on hydrogel-based materials (Weichao Li et al., 2022; Weizheng Li et al., 2022; Rattan et al., 2018; Scetta et al., 2021; Zhao et al., 2021) have demonstrated a decrease in effective stiffness due to microstructure breakdown (physical bonds, crosslinks, hard/soft segments) under fatigue loading. Considering cartilage as a hydrogel, the reduced apparent stiffness, non-linearity (Figure 5), and irreversible thickness recovery (Figure 4) could be attributed to overall damage to the structural integrity of the cartilage. The creep behavior observed in the fatigue loading (1 Hz and 3000 cycles) is aligned with compaction phenomenon observed in literature on cartilage (Gao et al., 2015; Kerin et al., 2003; Maier et al., 2019; Simon et al., 1990) and soft engineering materials (Lu et al., 2023; Ni et al., 2020; Zhao et al., 2021).

Overall, the decrease in phase shift and energy dissipation before and after compressive cyclic loading confirms changes in the dynamic behavior of cartilage. The experimental data partially support our hypothesis of a reduction in energy dissipative properties, however, because neither parameter changed with altered stress level, the results may suggest that loading frequency (Chawla et al., 2024) and cycle number would have a greater influence than stress level on these metrics. Reduced energy dissipation following cyclic (fatigue) loading reported in this study aligns with previous research that investigated decreased dynamic behavior in cartilage (Bellucci and Seedhom, 2002; Vazquez et al., 2019) and decreased energy dissipation in soft engineering materials (Ni et al., 2020; Scetta et al., 2021; Zhao et al., 2021). Experimental data in this work did not reveal a consistent trend in energy dissipation or phase angle with an increase in stress levels (Figure 6), differing from previous studies demonstrating more pronounced fatigue-induced damage at higher peak stress levels (Gao et al., 2015; Movahedi-Rad et al., 2018).

*4.3 Local damage after cyclic loading: Crack growth at different physiological stress levels*

The extension of cracks in cartilage under cyclic loading implies localized damage (Figure 8) in the vicinity of the microfractured area (damage zone), resulting in reduced fracture and fatigue resistance. Our experimental findings support our hypothesis, indicating that the change in crack length, $\Delta a$, increased with the applied maximum stress (Figure 7; Figure S4-S10). Contrary to crack propagation laws (Frost and Dugdale, 1958; Head, 1953; Liu, 1961; McEvily Jr and Illg, 1958; Paris and Erdogan, 1963) (highlighted in *section 4.1* above), this study has investigated localized crack growth as function of applied stress levels due to the likelihood of complex local stress states (decoupling mixed modes responsible for crack growth with experimental findings) in the vicinity of the damaged area. The authors predict existence of mode I crack extension (denoted by black arrows in Figure 8) due to tensile stresses orthogonal to the crack boundary onto the surrounding undamaged zone. Additionally, there could be a possible existence of mode II crack extension (denoted by white arrows in Figure 8) due to in-plane shear stresses.

*4.4 Limitations*

While this investigation reveals changes in bulk mechanical behavior and localized crack growth rate in cartilage as a function of stress level, it is not without limitations. The preparation of non-curled flat cartilage samples, the need to prevent minor structural-mechanical alterations during extended fatigue testing and obtaining consistent crack morphology (during crack initiation) are some of the experimental challenges. Limited higher range of the stress levels (although targeted physiologically-relevant stresses) that could be achieved with the testing setup was another constraint. Additionally, the study did not incorporate a model for local stress distribution in the damaged area to establish a correlation between global and local damage, which would allow a detailed estimation of failure modes. Similarly, we did not directly evaluate sources of energy dissipation, and sources such as adhesion between the sample and the platen and friction at the crack surfaces may confound the measurements of phase angle and dissipated energy. To address these limitations, the application of computational analyses such as cohesive zone modeling would help in understanding localized damage zones and crack propagation mechanics in the non-linear viscoelastic material like cartilage, yet, those analyses are outside of the scope of this study. Furthermore, quantifying microstructural changes post-fatigue loading could offer more substantial evidence regarding the extent of fatigue-induced damage at various stress levels. Lastly, employing techniques like digital image correlation or a high-speed camera setup to capture *in-situ* material response and crack growth behavior holds promise for generating insightful findings in future research.

## 5. Conclusions

This study explores the impact of physiological stress levels on the energy dissipative and crack growth behavior of articular cartilage subject to compressive cyclic (fatigue) loading. This study confirms that increasing stress levels (first Piola-Kirchhoff compressive stress) from light-weighted activities ($\sigma_{max}$ = 1.77 MPa) to strenuous activities ($\sigma_{max}$ = 10.61 MPa) results in greater changes in the structural integrity as evidenced by reduced thickness ($\overline{th}$) and reduced apparent stiffness ($\overline{C_1 C_2}$), indicating bulk cartilage damage. The overall reduction in dynamic properties, such as energy dissipation (*ED*) and phase angle ($\phi$) suggest fatigue-induced damage following cyclic loading. Crack growth as a function of increasing stress level indicates diminished fatigue-resistance as expected under higher stresses. The research sheds light on the connection between global damage (alterations in bulk mechanical behavior) and localized damage (crack growth) in articular cartilage under fatigue loading. This study enhances researchers' understanding of cyclic response of cartilage by investigating fundamental changes in both global and local material properties. Ultimately, the insights gained from this research could be valuable for clinicians in mitigating tissue degeneration during osteoarthritis and contribute to the development of bio-inspired soft materials for engineering applications.

**CRediT authorship contribution statement**
**Dipul Chawla:** Conceptualization, Methodology, Data curation, Validation, Formal analysis, Investigation, Writing- original draft, review & editing. **Eric Kazyak:** Data curation, Formal analysis, Writing- review & editing. **Melih Eriten:** Conceptualization, Formal analysis, funding acquisition, Writing-review & editing. **Corinne R. Henak:** Conceptualization, Formal analysis, Investigation, Project management, funding acquisition, Writing- original draft, review & editing.

**Declaration of Competing Interest**
The authors declare that they have no known competing financial interests or personal relationships that could have appeared to influence the work reported in this paper.

**Acknowledgments**
This research study is partially funded by the US National Science Foundation (award number: CMMI-DCSD-1662456, award number CMMI-BMMB-2225174) and University of Wisconsin, Madison: (VCGRE, William F. Vilas Trust). Tissue from UW-Madison Meat Lab is gratefully acknowledged.

**Supplementary Information for:**

# Influence of applied stress on energy dissipation and crack growth in articular cartilage

Dipul Chawla, Eric Kazyak, Melih Eriten, Corinne R. Henak

## Curve-fitting for global and local damage

Results from diagnostic testing before and after fatigue were used to determine the fatigue-induced damage (global) to the samples. Normalized output response was correlated with applied stress level using three laws; namely linear-fit, exponential fit, and power-law fit not shown in the main text are shown in Figure S1-S3. The power-law fit had the highest correlation coefficient, hence used in the main text.

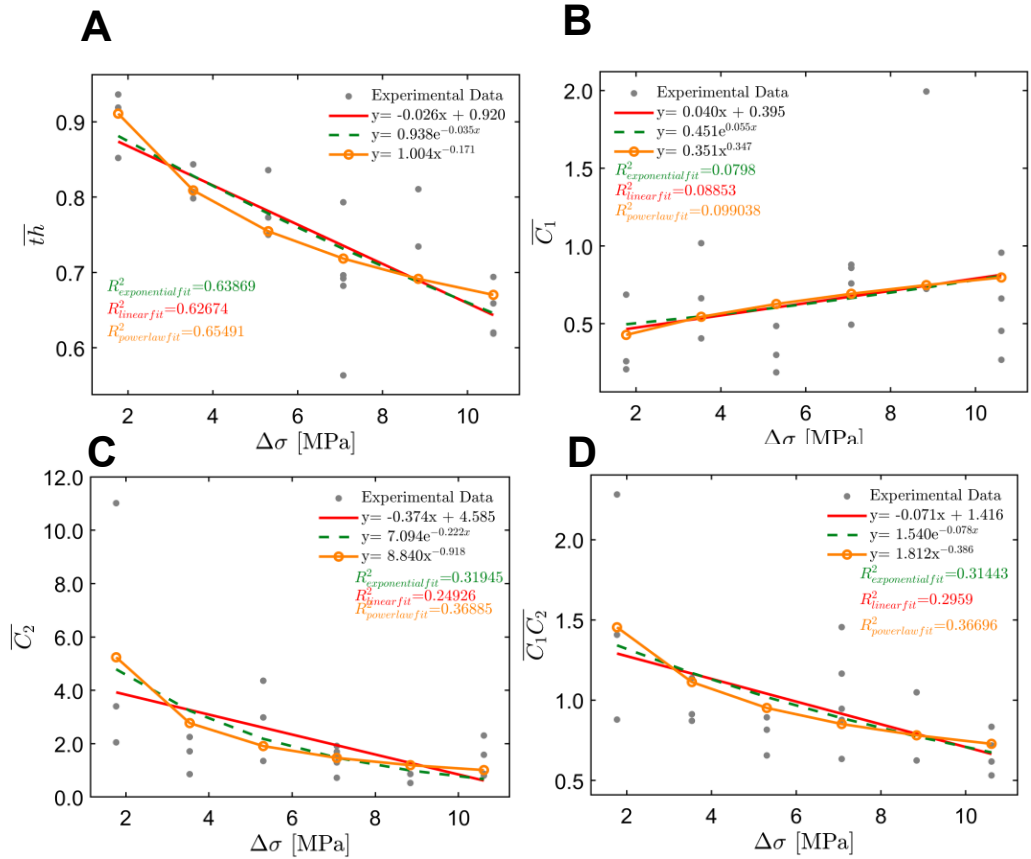

**Figure S1.** Effect of different physiological stress levels on structural integrity and mechanical properties using constitutive model (Veronda-Westmann). (A) Normalized thickness $\overline{th}$ decreased, (B) Normalized nonlinear first Veronda-Westmann material constant $\overline{C_1}$ unchanged (weak correlation), (C) Normalized nonlinear second Veronda-Westmann material constant $\overline{C_2}$ decreased, and (D)

Normalized apparent stiffness $\overline{C_1 C_2}$ after compressive cyclic loading with an increase in stress levels indicating cartilage softening and global structural damage. Red lines represent linear fit, green line represents exponential fit, and then the orange line represents power-law fit.

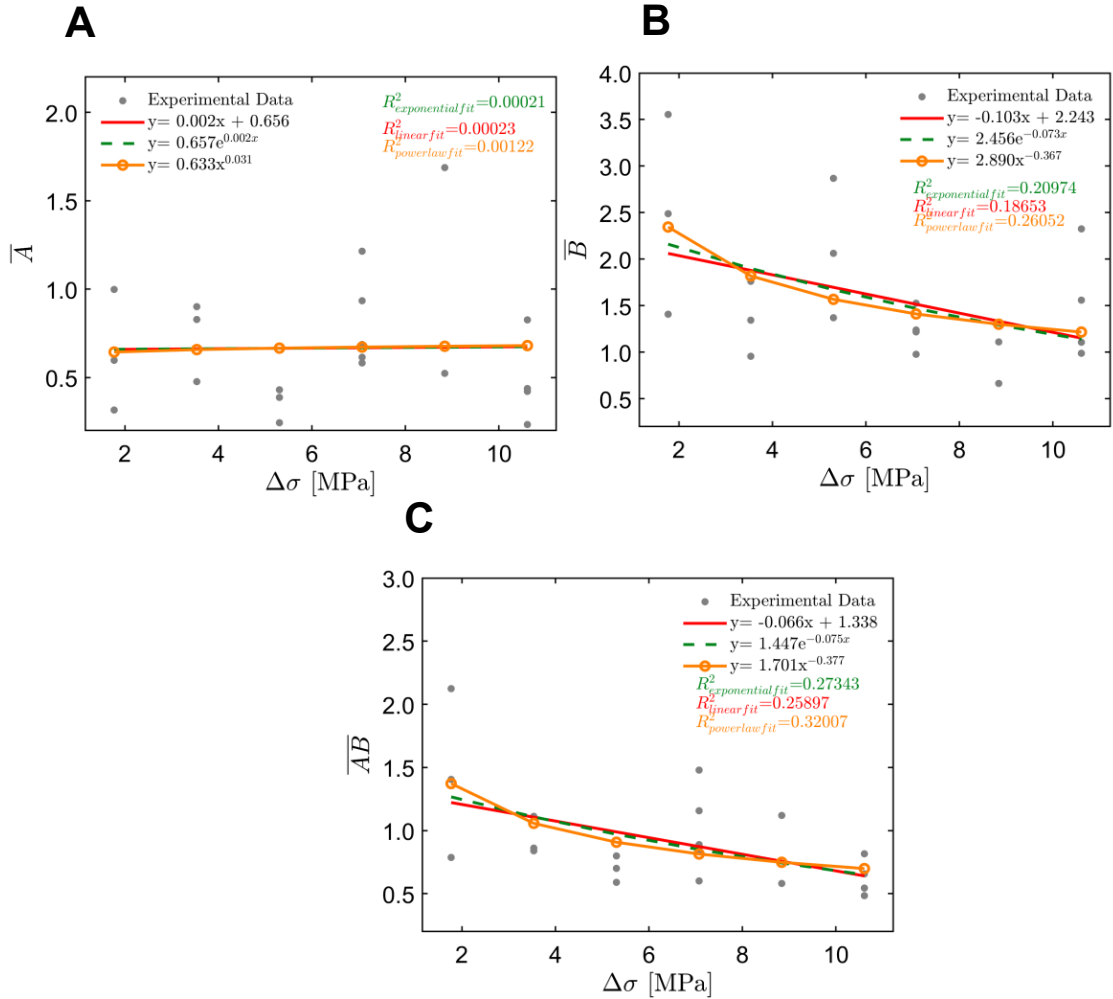

**Figure S2.** Effect of different physiological stress levels on structural integrity and mechanical properties using phenomenological fit; $\sigma = A(e^{B*\varepsilon} - 1)$. (A) Normalized first material constant $\bar{A}$ unchanged (weak correlation), (B) Normalized second material constant $\bar{B}$ decreased, and (C) Normalized apparent stiffness $\overline{AB}$ after compressive cyclic loading with an increase in stress levels indicating cartilage softening and global structural damage. Red lines represent linear fit, green line represents exponential fit, and then the orange line represents power-law fit.

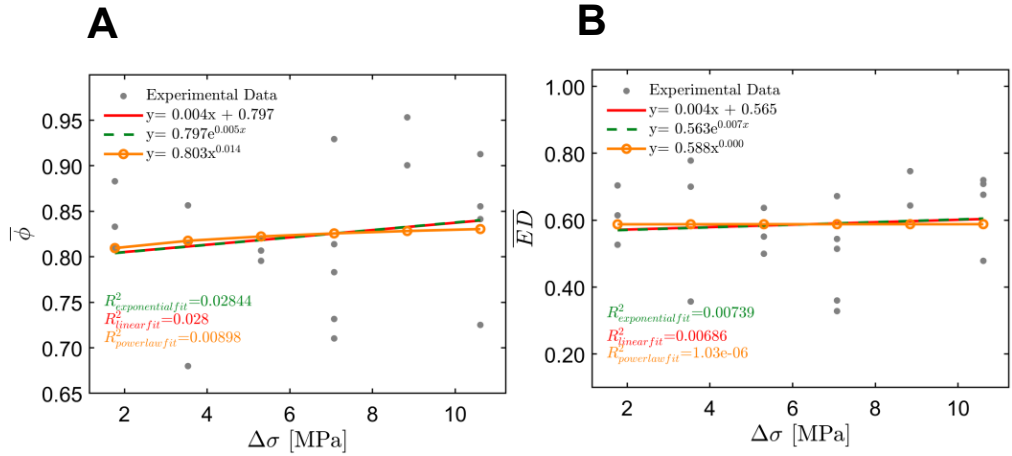

**Figure S3.** Effect of different physiological stress levels on dynamic mechanical properties under compressive cyclic loading. (A) Normalized phase angle $\bar{\phi}$ and (B) Normalized energy dissipation ratio $\overline{ED}$ were not correlated with physiological stress levels. Red lines represent linear fit, green line represents exponential fit, and then the orange line represents power-law fit.

## Crack Imaging and Crack Length Measurement

3D images of cracks were taken before and after cyclic loading using a LEXT laser microscope to measure crack lengths. Images were acquired at 10×. The normalized change in crack length, $\Delta a/a$, was also shown to follow a power law, $y = 0.050x^{0.308}$ and $R^2=0.32$, with increase in stress levels (Figure S4). All 2D overview images of the cracks are shows in Figures S5-S10.

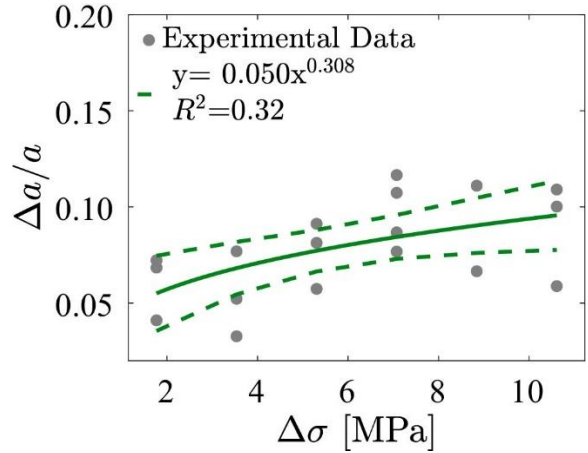

**Figure S4.** Normalized change in crack length; $\Delta a/a$ following a power with increase in physiological stress level ($\Delta\sigma$). Dashed lines show the 95% confidence intervals.

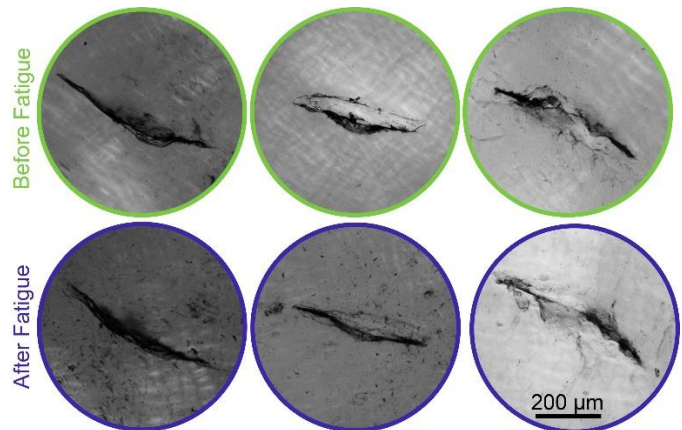

**Figure S5.** Crack images at L1 stress level, $\sigma_{max}$= 1.77 MPa. (top row) before fatigue (cyclic loading), and (bottom row) after fatigue (cyclic loading).

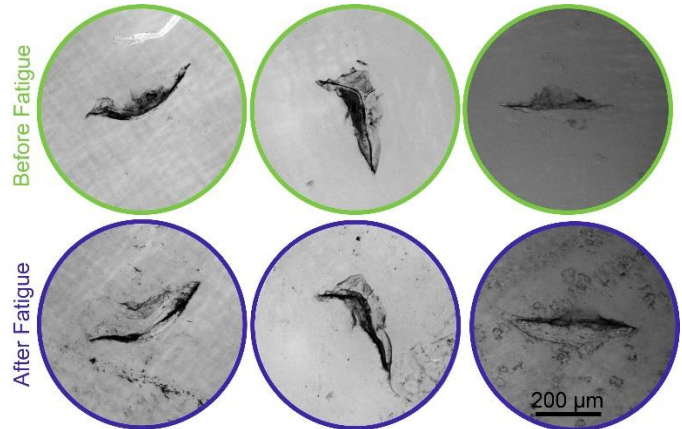

**Figure S6**. Crack images at L2 stress level, $\sigma_{max}$= 3.54 MPa. (top row) before fatigue (cyclic loading), and (bottom row) after fatigue (cyclic loading).

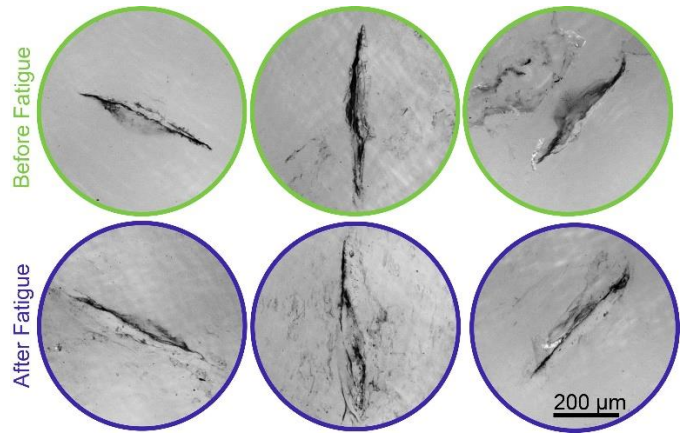

**Figure S7.** Crack images at L3 stress level, $\sigma_{max}$= 5.31 MPa. (top row) before fatigue (cyclic loading), and (bottom row) after fatigue (cyclic loading).

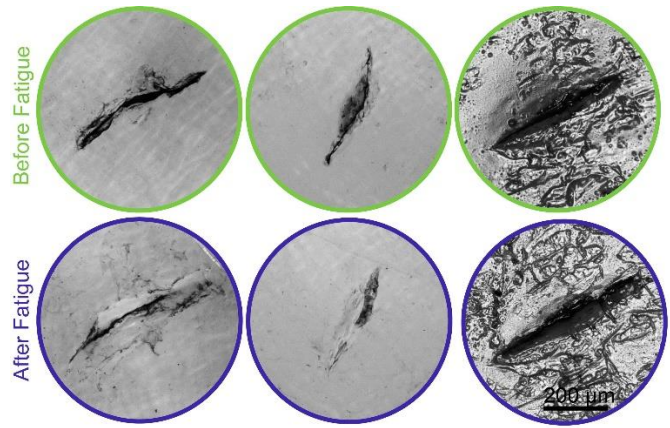

**Figure S8.** Crack images at L4 stress level, $\sigma_{max}$= 7.07 MPa. (*top row*) before fatigue (cyclic loading), and (*bottom row*) after fatigue (cyclic loading).

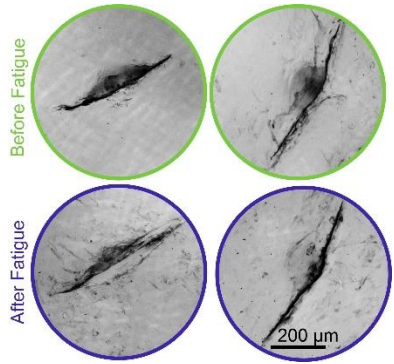

**Figure S9.** Crack images at L5 stress level, $\sigma_{max}$= 8.84 MPa. (*top row*) before fatigue (cyclic loading), and (*bottom row*) after fatigue (cyclic loading).

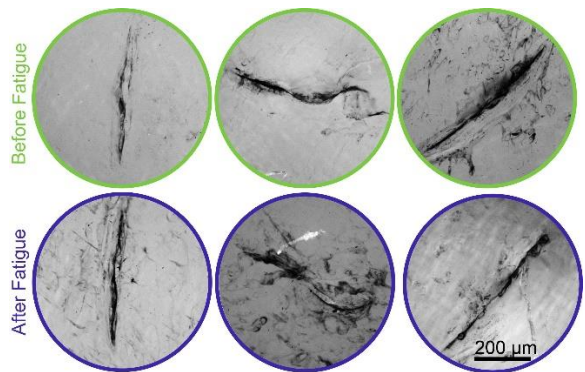

**Figure S10.** Crack images at L6 stress level, $\sigma_{max}$= 10.61 MPa. (*top row*) before fatigue (cyclic loading), and (*bottom row*) after fatigue (cyclic loading).

## Correlation between local damage ($\Delta a$) vs global damage ($\Delta thickness$)

3D images of cracks were taken before and after cyclic loading using LEXT laser microscope to measure crack lengths ($a$). The thickness of the sample before pre-diagnostics ($th$) was measured using digital caliper while the thickness of the sample before post-diagnostics ($th_{\text{post}}$) was evaluated from the controller position. Details of the measurement methods are in main text.

Change in the crack length (local damage around the crack tip) was correlated with the change in the thickness of the cartilage (global structural damage) using three laws; namely linear-fit, exponential fit, and power-law fit.

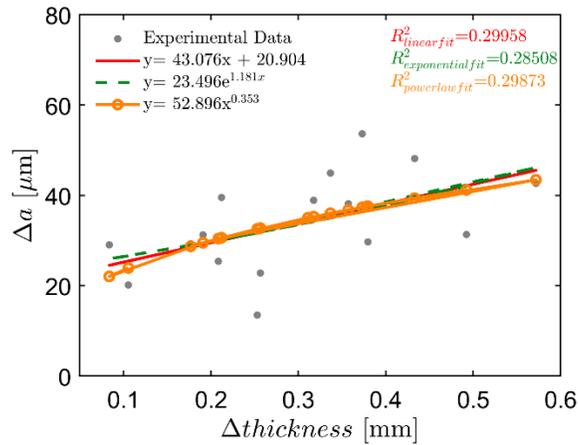

**Figure S11.** Change in crack length *vs* change in thickness of the tissue after compressive cyclic (fatigue) loading. Red lines represent linear fit, green line represents exponential fit, and then the orange line represents power-law fit.